\documentclass[preprint,12pt]{elsarticle}

\usepackage{amssymb}
\usepackage{amsmath}
\usepackage{comment}
\journal{Nuclear Physics B}

\begin{document}

\begin{frontmatter}



\title{Reduced Markovian Models of Dynamical Systems}


\author[inst1]{Ludovico Theo Giorgini}
\author[inst2]{Andre N. Souza}
\author[inst3]{Peter J. Schmid}

\affiliation[inst1]{organization={Nordita, Royal Institute of Technology and Stockholm University},%Department and Organization 
            city={Stockholm},
            country={Sweden}}

\affiliation[inst2]{organization={Massachusetts Institute of Technology},%Department and Organization 
            city={Cambridge}, 
            state={Massachusetts},
            country={United States}}
            
\affiliation[inst3]{organization={King Abdullah University of Science and Technology (KAUST)},%Department and Organization 
            city={Thuwal}, 
            country={Saudi Arabia}}

\begin{abstract}
Leveraging recent work on data-driven methods for constructing a finite state space Markov process from dynamical systems, we address two problems for obtaining further reduced statistical representations. The first problem is to extract the most salient reduced-order dynamics for a given timescale by using a modified clustering algorithm from network theory. The second problem is to provide an alternative construction for the infinitesimal generator of a Markov process that respects statistical features over a large range of timescales. We demonstrate the methodology on three low-dimensional dynamical systems with stochastic and chaotic dynamics. We then apply the method to two high-dimensional dynamical systems, the Kuramoto-Sivashinky equations and data sampled from fluid-flow experiments via Par\-ti\-cle-Image Velocimetry. We show that the methodology presented herein provides a robust reduced-order statistical representation of the underlying system.
\end{abstract}


% \begin{highlights}
% \end{highlights}


\end{frontmatter}



\section{Introduction}

Dynamical systems are ubiquitous in many scientific fields, ranging from physics and engineering to biology and finance \cite{te2020classical,peotta2021determination, ahmadi2020learning,ju2021fault}. The behavior of these systems is critical to our understanding of the world around us and our ability to forecast future events \cite{te2020classical, ahmadi2020learning, Lim2020Predicting, giorgini2020precursors, giorgini2021modeling}. One of the key challenges in studying dynamical systems is the detection of patterns, the reduction of their dimensionality, and the extraction of coherent structures that govern their intrinsic behavior \cite{ju2021fault, peotta2021determination, giorgini2022non, keyes2023stochastic, schmid2010dynamic, geogdzhayev2024evolving, souza2023statistical}. In this effort, clustering algorithms are commonly used as tools for identifying and categorizing patterns in data. In recent years, the application of clustering algorithms to dynamical systems has gained significant attention, as they provide a powerful way of studying the behavior of general systems and extracting meaningful insights into their structure \cite{Chiappa2008, Fernex2021, Hsu2020, souza2023transforming}. Moreover, when applied to time series data, they serve as partitions of state space, allowing for statistical reduced-order models of the underlying dynamics \cite{ChaosBook}.

This work addresses two separate issues:
\begin{enumerate}
    \item The construction of a minimal partition of state-space on a given time\-scale.
    \item The construction of an \textit{infinitesimal} generator that respects transition probabilities over a large range of timescales.
\end{enumerate}
By providing solutions to the above issues we can then construct a reduced-order statistical model of the underlying dynamical system.

The first problem is addressed through a modification of the Leicht-Newman algorithm \cite{Leicht2008}, which generalizes the algorithm from directed graphs to Perron-Frobenius operators. The original Leicht-Newman algorithm is based on modularity maximization which has a number of known problems, see \cite{PhysRevE.81.046106, Fortunato_2007, Fortunato_2016, Peixoto_2023}. However, we find that our modified modularity criteria, along with the application of the methodology to ``networks" arising from dynamical systems, yield results which garner insight to the original system. The new modularity criterion gives an automated way of determining a stopping criteria when performing spectral bisection as in \cite{Klus2022}, yielding a ``hands off" approach towards finding a reduced order system.

The second problem is addressed by using a perturbative approach to the construction of the generator. In essence, we use a preliminary algorithm to obtain an approximate generator, see \cite{souza2023statistical}. Then, using the Koopman eigenfunctions of this preliminary generator, we correct the eigenvalues of the preliminary generator by using the temporal autocorrelations of the Koopman eigenvectors. The new generator retains the eigenvectors of the original operator while using the modified eigenvalues.

The paper is divided into five subsequent sections. Section \ref{sec:math} reviews the mathematical setting for this study. Following this, Sections \ref{sec:mod_ln} and \ref{sec:perturb_generator} introduce the main algorithms: a modified Leicht-Newman algorithm and a perturbative construction of the infinitesimal generator, respectively. Section \ref{sec:method} summarizes the overall methodology. And finally, in Section \ref{sec:results}, we apply the methodology to a range of dynamical systems.

\section{Mathematical background}
\label{sec:math}
We consider the temporal evolution of an autonomous continuous-time  dynamical system of the general form
\begin{equation}
\dot{x}(t)=F(x(t)),
\label{x_cont}
\end{equation}
where $x:[0,T] \to \mathbb{R}^D,\, T>0$ represents the $D$-dimensional state of the system and $F:\mathbb{R}^D \to \mathbb{R}^D$ denotes a deterministic $D$-dimensional flow. Extending the formulation to non-autonomous dynamical systems can be accomplished in a straightforward way by augmenting the dimension of the state vector. Furthermore, the formulation can take into account multi-scale characteristics by rescaling each state-vector component according to a time-scale $\tau_i$ on which the respective dynamics are realized, i.e., we take $x_i(t/\tau_i) \to y_i(t)\; \forall i \in [1,D]$.

In many cases, the number of snapshots of the dynamical system and the degrees of freedom of its dynamics makes it impractical, or even infeasible, to study its behavior directly in the high-dimensional state space. Instead, a coarse-grained model is sought that preserves the statistical and dynamical features of the original full-scale system, and is more amenable to analytical tools. 

This coarse-grained model is constructed by defining a distance $d$ between the temporal snapshots, then utilizing this measure to cluster the snapshots into aggregates of similar states~\cite{Fernex2021}. In other words, snapshots close in state space are gathered into the same cluster, and the original time series is reduced to a low-dimensional analog for the inter-cluster dynamics. The choice of distance measure and the number of clusters $N$ is generally not defined $\textit{a priori}$ and may crucially depend on the details of the system under investigation and on the physical quantities of interest.  

Any dimensionality reduction introduces a stochastic component into our coarse-grained system that accounts for the inevitable information loss of the precise trajectory followed by the system in state space. Consequently, a probabilistic formalism to study the dynamics of the coarse-grained system is prudent and essential. More specifically, the temporal evolution of our coarse-grained system will be described by the following Markov process

\begin{equation}
X_{n+1} = S(X_n) = s(X_n)+W(X_n),
\label{X_cont}
\end{equation}
where $X:[0,T] \to \mathbb{R}^{N^m},\,T>0$ describes the state of the system, considering a $m$-dimensional delay embedding. The force fields $s:\mathbb{R}^{N^m} \to \mathbb{R}^{N^m}$ and $W:\mathbb{R}^{N^m} \to \mathbb{R}^{N^m}$ represent the deterministic and stochastic components, respectively. The amplitude of the stochastic force field $W$ can be reduced by increasing the values of $N$ and $m$. For our subsequent analysis, we will consider $m=1;$ a generalization of our results to arbitrary values of $m$ is straightforward.

The forward evolution of the probability distribution function (PDF) $\rho$ of the coarse-grained state $X_n$ follows a Fokker-Planck equation according to

\begin{equation}
\partial_t \rho = \mathcal{L}_{FP}\rho
\end{equation}
with $\mathcal{L}_{FP}$ as the Fokker-Planck operator. Its discrete counterpart can be stated as 

\begin{equation}
\rho_{t+\Delta t} = P^{\Delta t}\rho_t,
\label{rho_disc}
\end{equation}
where $P^{\Delta t}$ denotes the Perron-Frobenius operator or the transition matrix for a time step $\Delta t$.

We start by partitioning the state space by using k-means clustering of the snapshots of the system according to their Euclidean state-space distance. This preliminary partitioning acts as a starting point for constructing strongly intra-connected regions. These communities of strongly intra-connected elements are referred to as almost-invariant sets in the pertinent literature~\cite{Froyland2003}. A partitioning of state space into almost-invariant sets $\{\hat{A}_1^n,\hat{A}_2^n,\dots,\hat{A}_k^n\}$ satisfies
\begin{equation}
R_n(\hat{A}_1^n,\hat{A}_2^n,\dots,\hat{A}_k^n) = \textrm{sup}\{R_n(A_1,A_2,\dots,A_k)\},
\end{equation}
with $\{A_1,A_2,\dots,A_k\}$ as a measurable partition of $X$ and $R_n$ defined as 
\begin{equation}
R_n(A_1,A_2,\dots,A_k) = \frac{1}{k}\sum_{l=1}^k \frac{m(A_l\cap S^{-n}(A_l))}{m(A_l)},
\label{eq:Rn}
\end{equation}
where $m(B)$ denotes a Lebesgue measure of $B,$ and $n$ represents the number of time steps considered in the definition of the almost-invariant set. The operator $S^{-n}$ stands for the backstep operator over $n$ timesteps.

\section{A Modified Leicht-Newman Algorithm for Perron-Frobenius Operators}
\label{sec:mod_ln}
In what follows, we propose a modification of the directional Leicht-Newman algorithm~\cite{Leicht2008}, which, in its original form, consists of a recursive division of a network of size $N$ into two disjoint communities. Each network vertex $i$ is labeled by $s_i \in \{-1,1\}$ depending on which community it has been assigned to, and the values of $s_i$ are chosen to minimize the modularity parameter $\mathcal{M}$ defined as 
\begin{equation}
\mathcal{M} = \frac{1}{2N}\sum_{ij}s_i (B_{ij}+B_{ji}) s_j,
\end{equation}
with $B_{ij}$ as the modularity matrix given as 
\begin{equation}
B_{ij} = \frac{1}{N}\left(A_{ij} - \frac{k_i^{in}k_j^{out}}{N} \right).
\end{equation}
In the latter expression, $A$ stands for the graph adjacency matrix, and $k_i^{in},k_j^{out}$ represent the in- and out-degrees of each vertex, respectively.

Following~\cite{Newman2006}, the label vector $s$ can be written as a linear combination of the eigenvectors $v_i$ of $B+B^T$, i.e., $s=\sum_i a_i v_i$, with $a_i = v_i^T s$.
The modularity parameter then becomes
\begin{equation}
\mathcal{M} =\sum_i a_i v_i^T(B+B^T)\sum_j a_j v_j = \sum_i \beta_i(v_i^T s)^2,
\end{equation}
with $\beta_i$ as the $i^\textrm{th}$ largest eigenvalue. $\mathcal{M}$ is maximized by taking $s$ maximally collinear to the principal eigenfunction $v_1$ (corresponding to the largest eigenvalue), with the constraint $s_i \in \{-1,1\}.$ This latter constraint is satisfied by choosing $s_i=\textrm{sign}\left(v_{1,i}\right)$.

In an effort to progressively divide the network into further communities, we continue to maximize the modularity parameter after replacing the original modularity matrix with the generalized modularity matrix
\begin{equation}
B^{g}_{ij}=B_{ij}-\delta_{ij}\sum_{k\in g}B_{ik},
\end{equation}
with indices $i,j$ ranging over the elements of the subgraph $g$ that we wish to further decompose into two sub-communities.

\subsection{The Modification}

In our case, we consider a network with $N$ vertices of out-degree $m$ (the in-degree is not specified). Each vertex represents a transition probability of $1/m$. We can write
\begin{equation}\begin{split}
B_{ij}^t &= \frac{1}{m N}\left( (\#_{ij})^t-\frac{(\#_{i}^{out})^t(\#_{j}^{in})^t}{m N}\right) \\ &=\frac{1}{N}\left( \frac{(\#_{ij})^t}{m}-\frac{1}{N}\frac{(\#_{i}^{out})^t}{m}\frac{(\#_{j}^{in})^t}{m}\right),
\end{split}\end{equation}
with $(\#_{ij})^t$ denoting the number of edges pointing from cluster $i$ to cluster $j$ at time $t,$ and $(\#_i^{in})^t$ ($(\#_{i}^{out})^t$) stands for the number of edges pointing towards (away from) the $i^{\textrm{th}}$ vertex. This calculation motivates the definition
\begin{equation}
B_{ij}^t = P_{ij}^t - \frac{1-P_{ii}^t}{N}\left(\sum_k P_{kj}^t \right).
\end{equation}
We then have to minimize the modularity parameter $\mathcal{M},$ using the modularity matrix defined above, until it falls below a user-supplied threshold $\mathcal{M}_{\textrm{min}}$. By increasing this threshold, we influence the intra-connectivity of the almost-invariant sets that the algorithm determines.

Before proceeding, we notice that the expression for the modularity matrix can be simplified for large values of $N$ and $t$ according to
\begin{equation}\begin{split}
B_{ij}^t &= P_{ij}^t - \frac{1-P_{ii}^t}{N}\left(\sum_k P_{kj}^t\right)\\ &= P_{ij}^t - \frac{1}{N}\left(\sum_{k\neq i}P_{ik}^t\right)\left(\sum_k P_{kj}^t\right) \\ &\simeq P_{ij}^t - P_{ij}^\infty,
\label{eq:mod_app}
\end{split}\end{equation}
where in the last step we used the fact that the sum of the rows or columns of the transition matrix converges faster to its asymptotic values than the single matrix elements.

\section{A Perturbative Construction of the Generator}
\label{sec:perturb_generator}
We can then state the explicit dependence of the Perron-Frobenius operator on the time variable. In the limit of $\Delta t \to 0$, Eq. (\ref{rho_disc}) becomes

\begin{equation}
\lim_{\Delta t \to 0} \rho_{t+\Delta t} = (I + Q \Delta t) \rho_t, 
\end{equation}
and, taking the continuous limit, we arrive at 

\begin{equation}
\dot{\rho} = Q \rho,
\end{equation}
that is, a forward evolution equation for $\rho$ -- discrete in space and continuous in time -- which can be formally solved to read 
\begin{equation}
\rho_{t} = e^{Q(t-s)} \rho_s = P^{t-s} \rho_s,
\label{PQequation}
\end{equation}
with $t>s$.

The matrix $Q$ is constructed from the coarse-grained data by noticing that, under the assumption of the Markov property for our system, the residence times in each cluster are independent and exponentially distributed with rates $r_i=1/\tau_i \; \forall i \in [1,N]$ with $\tau_i$ representing the mean residence time in cluster $i$. Therefore, we can write \cite{Klus2016}
\begin{equation}\begin{split}
\lim_{\Delta t \to 0} \frac{1-P_{ii}^{\Delta t}}{\Delta t} &= \lim_{\Delta t \to 0} \frac{\textrm{Pr}(\tau_i<\Delta t)}{\Delta t} = r_i = -Q_{ii},\\
\lim_{\Delta t \to 0} \frac{P_{ij}^{\Delta t}}{\Delta t} &= \lim_{\Delta t \to 0} \frac{1-P_{ii}^{\Delta t}}{\Delta t} \frac{P_{ij}^{\Delta t}}{\sum_{j\neq i} P_{ij}^{\Delta t}} \\ &= r_i \frac{P_{ij}^{\Delta t}}{\sum_{j\neq i} P_{ij}^{\Delta t}} = Q_{ij}.
\label{eq_QQ}
\end{split}\end{equation}

The fact that we constructed the matrix $Q$ using the statistics of the system at infinitesimal time scales, often introduces errors that become prominent when considering the statistical properties of the system at larger time scales. This is especially the case when the number of clusters is small. In what follows, we propose a method to overcome this issue by correcting the values of $Q$ using information about the system's behavior on larger time scales. These corrective terms are based on a matrix perturbation approach.

\subsection{The Perturbation Correction to the Generator}
Let us assume that the matrix $Q_{\textrm{pert}}$ approximates the matrix $Q$ that we constructed above, which means that we obtain $Q_{\textrm{pert}}$ from an additive perturbation to $Q.$ We write $Q_{\textrm{pert}}-Q=\delta Q = g Q'$ with $Q_{ij}' = {\mathcal{O}}(1)$ and $g \ll 1.$ The eigenvalue problem for $Q_{\textrm{pert}}$ can be stated according to 
\begin{equation}
(Q+g Q') (\phi_0^i + g\phi_1^i) = (\lambda_0^i + g\lambda_1^i)(\phi_0^i + g\phi_1^i),
\end{equation}
which yields, considering only linear ${\mathcal{O}}(g)$ terms, 
\begin{equation}
Q' \phi_0^i + Q\phi_1^i = \lambda_0^i \phi_1^i + \lambda_1^i\phi_0^i.
\label{g_first}
\end{equation}
Multiplying on the left by the transpose of the unperturbed left eigenfunction of $Q$, denoted by $(\psi_0^i)^T$, we obtain
\begin{equation}
Q' \phi_0^i= \lambda_1^i\phi_0^i,
\end{equation}
where we have used the fact that the unperturbed left eigenfunctions of $Q$ are bi-orthogonal to the unperturbed right eigenfunctions. Consequently, since the unperturbed right eigenfunction can be expressed as a linear combination of the unperturbed right eigenfunctions, the second term on the left-hand side of Eq. (\ref{g_first}) cancels the first term on the right-hand side after left-multiplication by $(\psi_0^i)^T$.
We hence recast the perturbation $\delta Q$ as a function of the unperturbed eigenfunctions and eigenvalues of $Q$ according to  
\begin{equation}
\delta Q = \sum_{i} \delta\lambda^i\phi_0^i(\psi_0^i)^T,
\end{equation}
with $\delta\lambda^i = g \lambda_1^i$.

In order to obtain $\delta\lambda^i$ we construct a $N$-dimensional time series from the coarse-grained system by associating with each value $X_n$ a $N$-dimensional vector containing, in each element, the $X_n$-th value of the unperturbed left eigenfunction of $Q.$ In other words, we perform the map $X_n \to \psi_{X_n}^i = Y_n^i\; \forall i$, where we omitted the subscript ${}_0$ for $\psi^i$. 

The correlation function for $\tilde{Y}^i=Y^i-\langle Y^i \rangle$ then becomes
\begin{equation}\begin{split}
& \mathcal{C}_{\tilde{Y}^i}(t) = \frac{(\psi^i)^T \textrm{diag}(\phi_1)[(\psi^i)^T (P^t-\textrm{diag}(\phi_1))]^T}{(\psi^i)^T \textrm{diag}(\phi_1)\psi^i}\\&=\frac{(\psi^i)^T \textrm{diag}(\phi_1)\left[(\psi^i)^T \left(\sum_{k\neq 1}e^{(\lambda_0^k+g\lambda_1^k) t}\phi^k(\psi^k)^T \right) \right]^T}{(\psi^i)^T \textrm{diag}(\phi_1)\psi^i} \\&=\frac{e^{(\lambda_0^i+g\lambda_1^i) t}(\psi^i)^T \textrm{diag}(\phi_1)\psi^i}{(\psi^i)^T \textrm{diag}(\phi_1)\psi^i}=e^{(\lambda_0^i+g\lambda_1^i) t}.
\end{split}\end{equation}
We determine this correlation function from data for each value of $i$ and obtain $\lambda_0^i+g\lambda_1^i$ from a least-squares fit. After computing the difference between each exponent and the unperturbed eigenvalue $\lambda_0^i,$ we then estimate the eigenvalue deviation $\delta \lambda^i$ and subsequently the corrective matrix perturbation $\delta Q$.

The correlation function of the coarse-grained time series $\tilde{X} = X - \langle X \rangle$ becomes
\begin{equation}
\mathcal{C}_C(t) =\frac{C^T \textrm{diag}(\phi_1)\left[C^T \left(\sum_{k\neq 1}e^{\lambda_k t}\phi^k(\psi^k)^T \right) \right]^T}{C^T \textrm{diag}(\phi_1)C},
\end{equation}
where $C$ is a vector containing the centers of the clusters, and  $\lambda_k$ represents the $k$-th eigenvalue of $Q_{\textrm{true}}$.

Succinctly, the proposed modification to $Q$ is to retain the eigenvectors of $Q$ but modify the eigenvalues by performing an exponential fit to the autocorrelation of the Koopman modes. We will demonstrate in Section \ref{sec:results} a marked improvement in using $Q_{\textrm{pert}}$ to construct the transition matrix with respect to $Q$.

\subsection{Connecting the Generator to Modularity Maximimization}
Writing the Perron-Frobenius operator as a function of the eigenvalues and eigenfunctions of the transition rate matrix $Q$ and using the approximation in Eq. (\ref{eq:mod_app}), we restate the modularity matrix and the generalized modularity matrix as
\begin{equation}
B^t_{ij} \simeq \sum_{k\neq 1}e^{\lambda_k t}\phi^k_i\psi^k_j,
\label{eq:B_eigs}
\end{equation}

\begin{equation}
(B^t_{ij})^g \simeq \sum_{k\neq 1}e^{\lambda_k t}\left[\phi^k_i\psi^k_j - \delta_{ij}\sum_{h\in g}\phi^k_i\psi^k_h\right],
\label{eq:Bg_eigs}
\end{equation}
with $i,j\in g$.

We notice from Eqs. (\ref{eq:B_eigs}, \ref{eq:Bg_eigs}) that all eigenvalues of the modularity and generalized modularity matrix exponentially tend to zero in the limit $t\to \infty$ and, consequently, the division of phase space into almost-invariant sets becomes impracticable.  No almost-invariant sets are found for large values of time since the elements of the transition matrix approach the equilibrium probability distribution function and 
\begin{equation}
\lim_{n\to \infty}  \frac{m(A\cap S^{-n}(A))}{m(A)} = 1.
\end{equation}
As a consequence, Eq. (\ref{eq:Rn}) will cease to depend on the choice of $\{A_1,A_2,\dots,A_k\}$ and a partition of the state space into almost-invariant sets is no longer defined.

\section{Methodology}
\label{sec:method}
In summary, we have presented two different algorithms. One algorithm determines almost-invariant sets in state space for a given time scale starting from an initial clusterization which ignores details of the system under study, see Section \ref{sec:mod_ln}. The other algorithm, in contrast, constructs the transition rate matrix, i.e., the transition matrix’s generator, which accurately recovers the latter for any time step, see \ref{sec:perturb_generator}. The two algorithms can be combined: determining almost-invariant sets based on the former algorithm, then using these sets in the latter to construct the transition rate matrix. 

We thus propose a four-step approach to cluster time series of a multi-dimensional dynamical system:
\begin{enumerate}
\item We apply a k-means algorithm to precluster the time series with a user-specified value of $k,$ the number of clusters. This step ensures that we capture the coarse-grained system dynamics and, at the same time, that each cluster contains a sufficient number of data points.
\item We next employ the modified Leicht-Newman algorithm to identify the almost-invariant sets of the dynamical system associated with its first $n$ dominant time scales. These time scales are determined by the inverse of the first $n$ largest eigenvalues of the generator of the Perron-Frobenius operator built from the k-means-clustered time series. We iterate the algorithm until the modularity parameter becomes negative, while we record its value at each subdivision.
\item We determine the optimal number of clusters $m$ for each time scale based on these values. A value of $m$ is chosen to only keep the most connected subgroups of fine clusters, which take on the highest modularity values, and to discard the less connected, and therefore less relevant clusters. Based on these values, we obtain the almost-invariant sets with the desired number of clusters for each time scale and the parameter $\Delta C.$
\item We finally apply the matrix-perturbative method described above to compute the generator of the transition matrix.
\end{enumerate}

After the first three steps, it is instructive to determine the robustness of each coarse clusterization to a variation of the corresponding time parameter, assessing changes in the unstable regions in state space at the interface between different almost-invariant sets. To this end, we execute the modified Leicht-Newman algorithm two additional times and vary the time parameter $t$, corresponding to the time scale under consideration, by $\Delta t \ll t$ while keeping $\mathcal{M}_{\textrm{min}}$ fixed. We hence obtain three different clusterizations for each value of the time parameter, $\{A_1^{t},A_2^{t},\dots,A_{m}^{t}\}$, $\{A_1^{t+\Delta t},A_2^{t+\Delta t},\dots,A_{m}^{t+\Delta t}\}$, $\{A_1^{t-\Delta t},A_2^{t-\Delta t},\dots,A_{m}^{t-\Delta t}\}.$ These sets of clusters are used to quantify changes in the assignment of fine clusters to coarse ones $\Delta C,$ initiated by a variation in the time parameter. We define 
\begin{equation}
\Delta C = \frac{\max\{\#(A_i^{t+\Delta t} \cap A_j^{t})\}+\max\{\#(A_i^{t-\Delta t} \cap A_j^{t})\}}{2 N},
\label{eq:GCC}
\end{equation}
with $i,j \in [1,m]$. That is, we define the quantity $\Delta C$ as the average of the number of intersections between the coarse clusters obtained at time $t+\Delta t$ and $t$, and between those obtained at $t-\Delta t$ and $t$ divided by the total number of fine clusters. Since each label of a coarse cluster is arbitrary, the clusters are chosen to maximize the number of intersections.

Since we slightly vary the time parameter in each cluster, we expect clusters belonging to the same almost-invariant set to maintain their assignment in each clusterization. In contrast, clusters at the interface will be sensitive to variations in the time parameter and will thus be assigned to different almost-invariant sets. The intersection of these clusterizations will then delineate the former, characterized by extensive regions of state space with a slow relaxation of the corresponding diagonal elements of the transition matrix, from the latter, formed by small regions at the interface of the former and distinguished by rapid escapes in phase space, associated with a fast decay of the associated diagonal elements of the transition matrix.

\section{Results} 
\label{sec:results}
We first apply the algorithms described above to three idealized models, whose dynamics are expressed entirely in three-dimensional state space and for which we assess the accuracy of our algorithms. Following this, we study data sets related to more complex and high-dimensional dynamical systems.

\begin{figure*}
 	\centering	\includegraphics[width=0.75\textwidth]{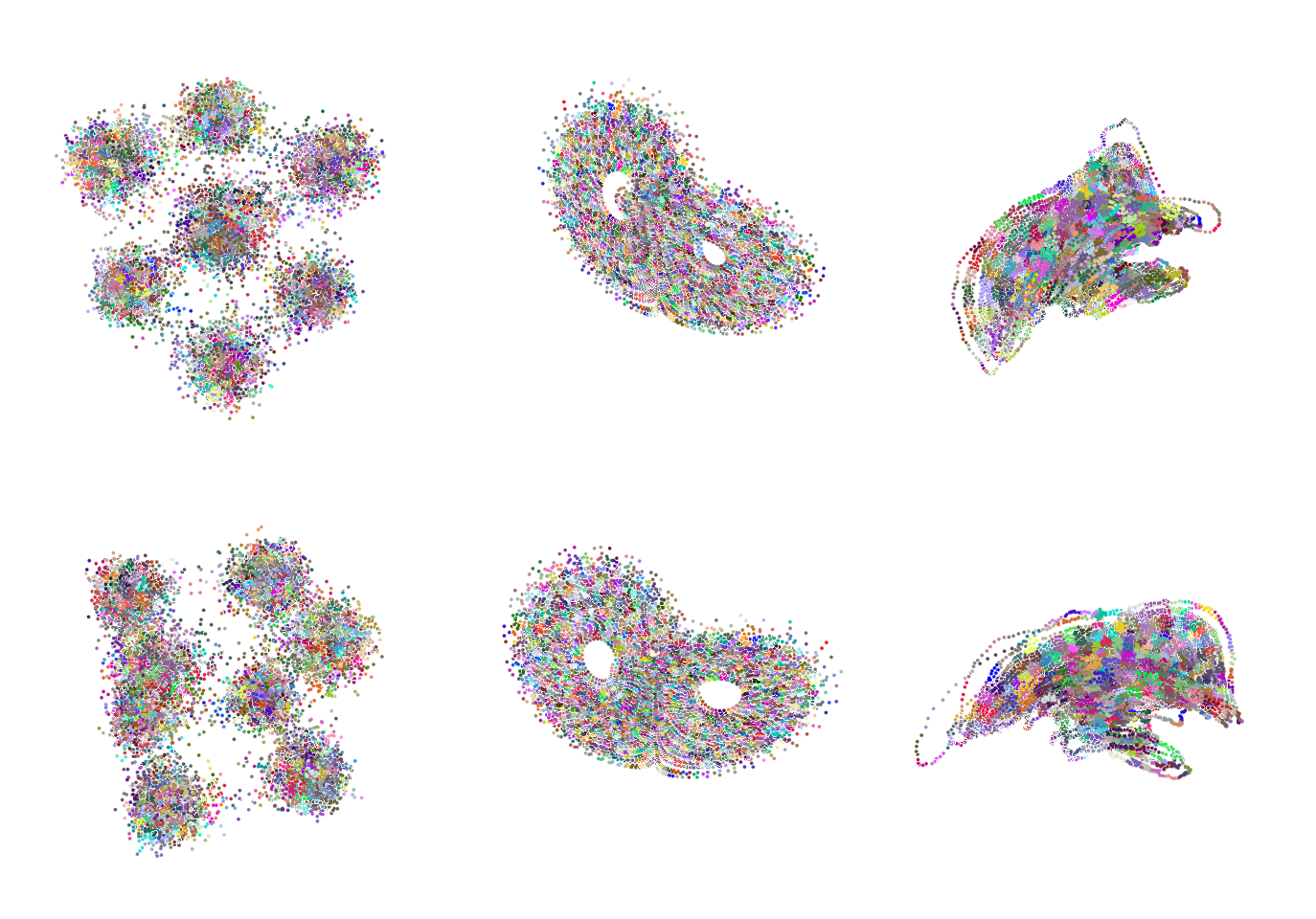}
 	\caption{Attractors for the three idealized models from different angles at the finest cluster resolution. The potential consists of eight minima in three-dimensional space with stochastic transitions between the wells. The Lorenz equations are a well-studied deterministic chaotic dynamical system. And the Newton-Leipnik system is augmented with noise and exhibits stochastic transitions between two chaotic attractors.}
 	\label{fig:three_systems}
\end{figure*} 

\begin{figure*}
 	\centering	\includegraphics[width=0.75\textwidth]{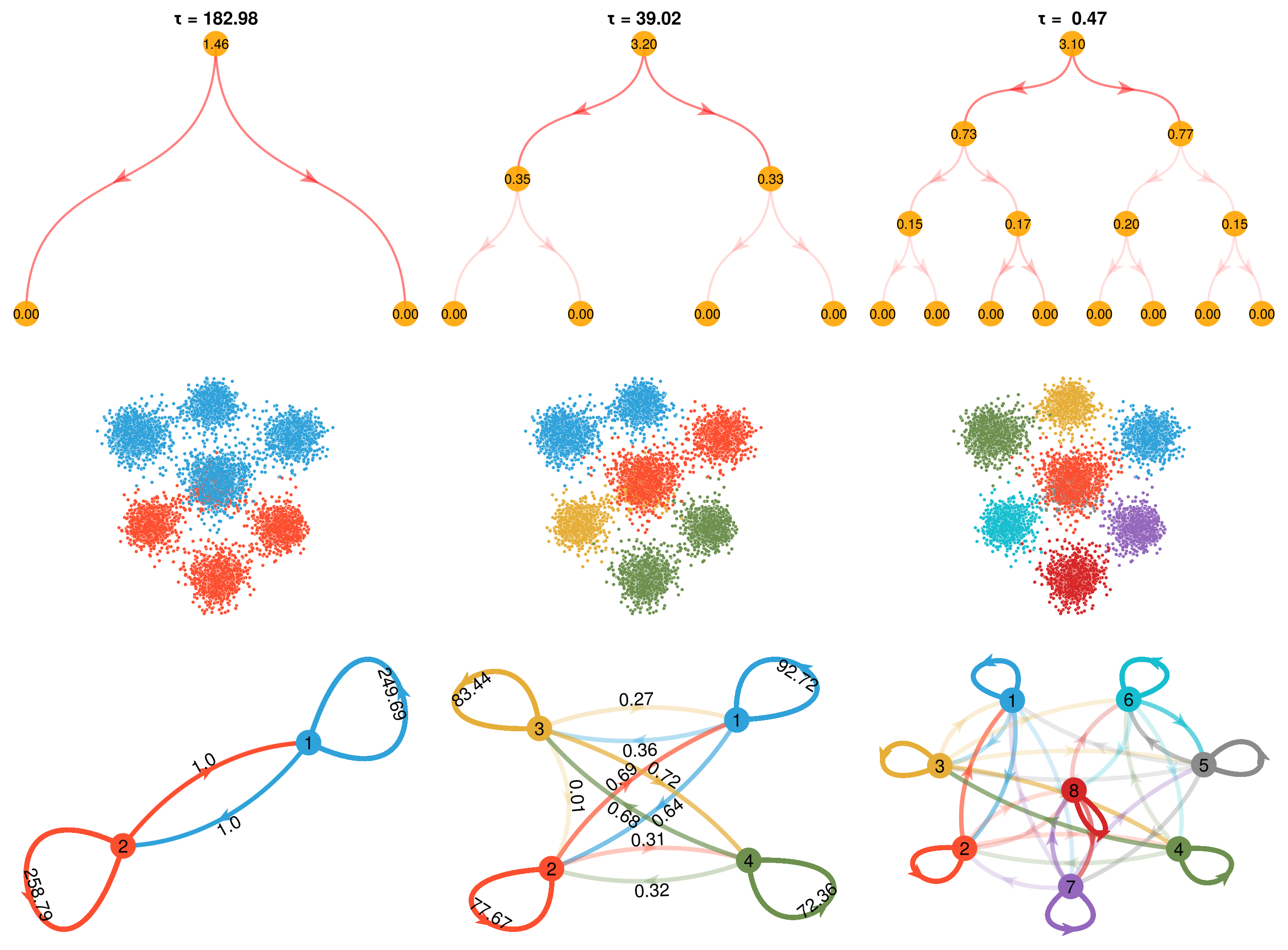}
 	\caption{The Modified Leicht-Newman Algorithm applied to the potential-well problem at various time scales. As the time scale becomes smaller, the algorithm distinguishes between more subsets of state-space, up to the finest resolution consisting of eight different states. The top row shows the binary partitioning based on the modified Leicht-Newman algorithm. The number in the circles indicates the modularity parameter at each classification stage of state space. The middle panel shows the classification of state space by coloring like states with similar colors. The last row shows the resulting graph structure of the partitioned state-space. The errors are $\Delta C = 0$ for the first two timescales and $\Delta C = 5.5 \times 10^{-5}$ for the last timescale.}
 	\label{fig:multiwell}
\end{figure*} 

\begin{figure*}
 	\centering	\includegraphics[width=0.75\textwidth]{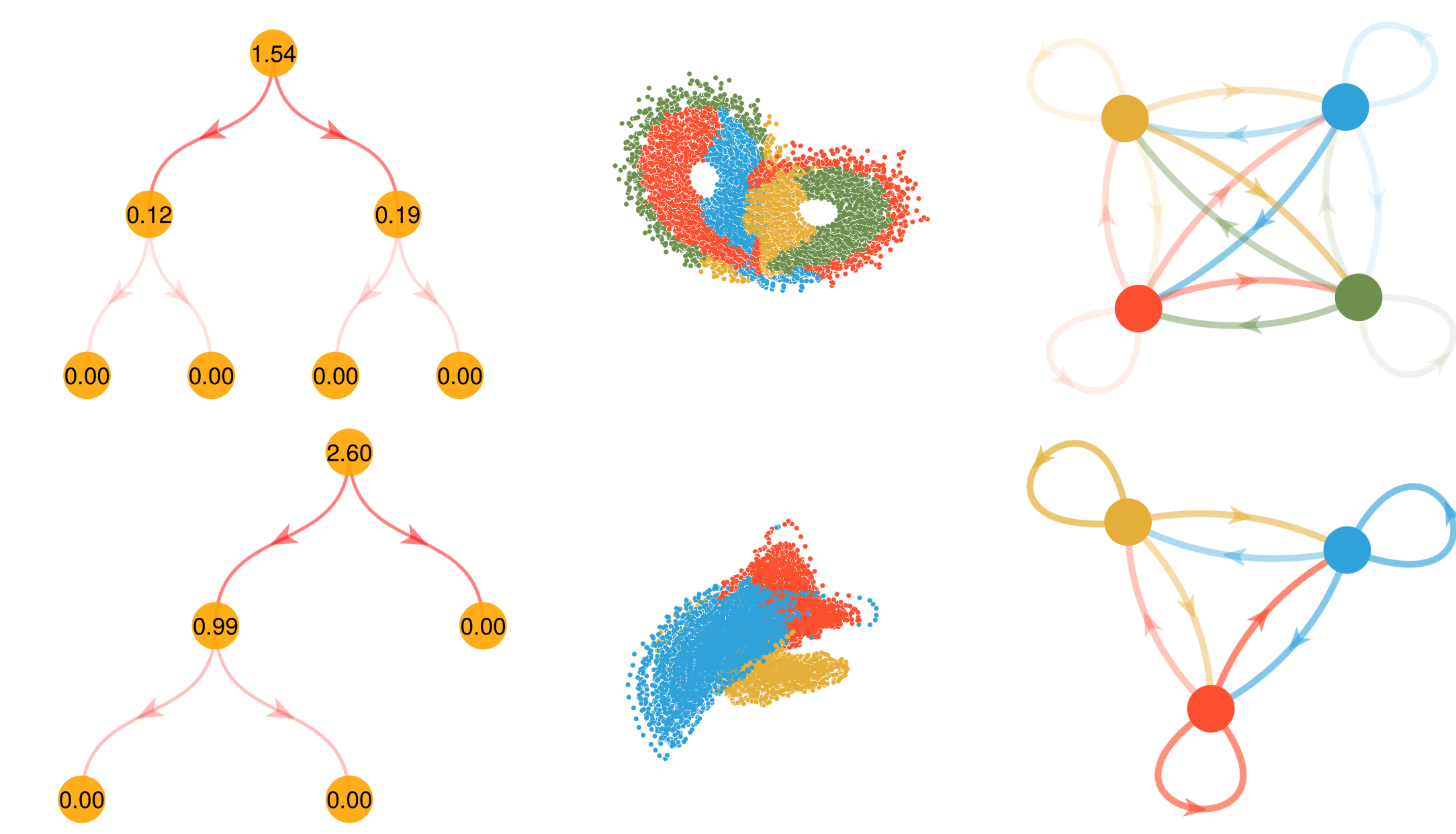}
 	\caption{The clustering algorithm applied to the Lorenz equations and the Newton-Leipnik system. We applied the modified Leicht-Newmann algorithm using a time scale $\tau \approx 0.63$  to the Lorenz system and a time scale $\tau \approx 0.89$ to the Newton-Leipnik system. These time scales correspond to the third non-zero eigenvalues of both systems. The Lorenz dynamics are partitioned into four subsets of state space. The Newton-Leipnik system results in a partitioning into three subsets of state-space, with the yellow region representing a separate attractor and the red-blue region representing a further partitioning of the larger chaotic attractor. The last column shows a representation of the transition probabilities between different subsets of state space. The score for the Lorenz system is  $\Delta C \approx 0.01683$ and for the Newton-Leipnik system is $\Delta C \approx 0.0002$.}
 	\label{fig:lorenznewton}
\end{figure*} 

\begin{figure*}
 	\centering	\includegraphics[width=1.\textwidth]{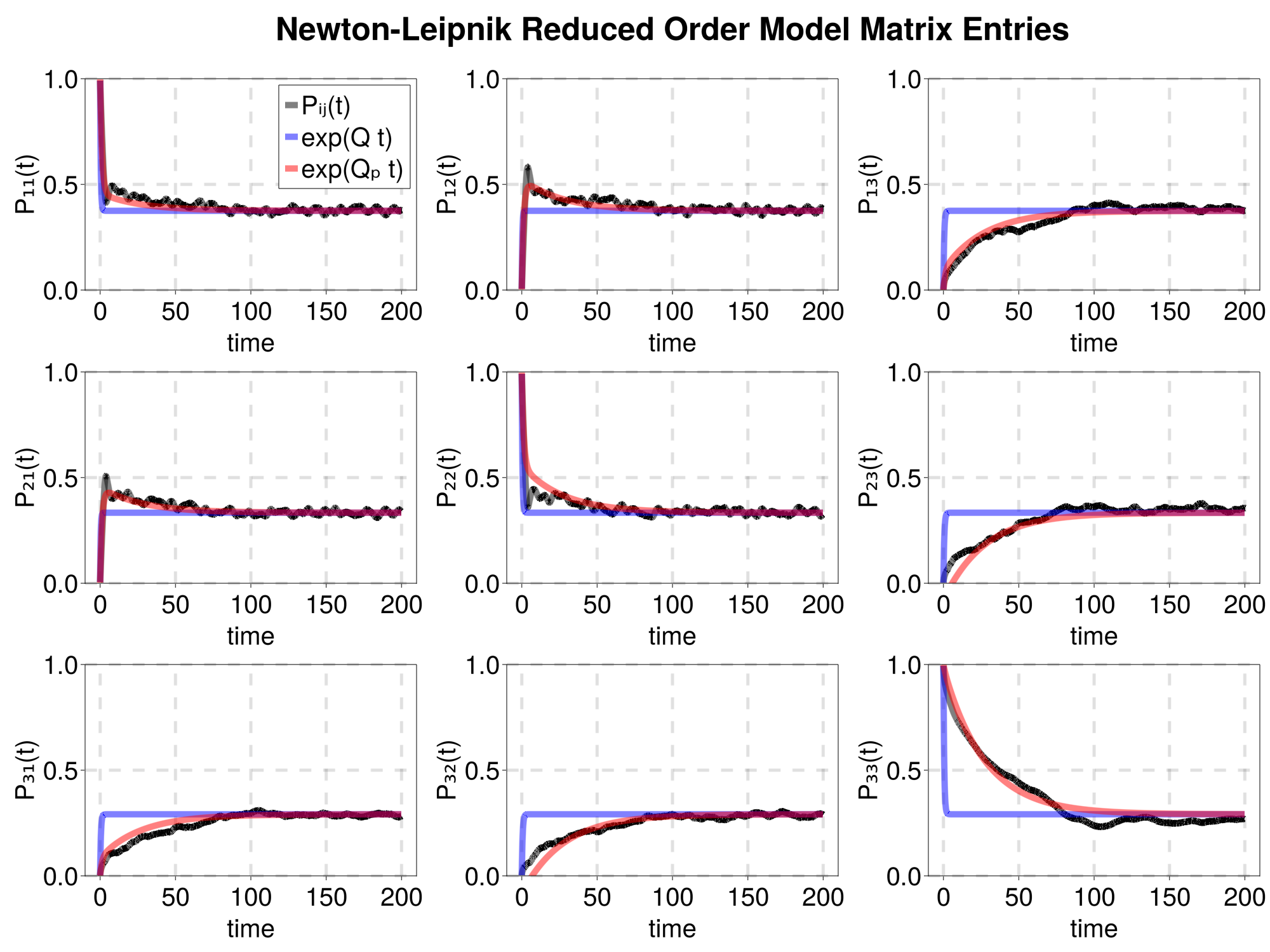}
 	\caption{Matrix elements of the transition matrix $P_{ij}(t)$ as a function of time, considering the almost-invariant sets of the Newton-Leipnik system corresponding to a time scale $\tau = 1.2$, the same used in Figure \ref{fig:lorenznewton}. The matrix elements have been constructed considering the time series of the Markov states (black curves), using $Q$ from Eq. (\ref{eq_QQ}) (blue curve) and $Q_{p}$ from our proposed method (red curves).}
 	\label{fig:pij}
\end{figure*} 

\subsection{Idealized Models}
Below we present the three idealized models for our preliminary test of the algorithms. In all three cases a k-means algorithm with 2000 centroids has been used to construct the fine clusters which will subsequently be used to determine the almost-invariant sets. See Figure \ref{fig:three_systems}.
\subsubsection{Three-dimensional potential well}
We consider a three-dimensional dynamical system whose state evolves in time according to the overdamped Langevin equation
\begin{equation}
\dot{x}(t)=-\nabla U(x) + \sigma \xi(t),
\label{Lang}
\end{equation}
with $t\in[0,3.3 \cdot 10^5]$, a time step $dt=0.03,$ and $U$ denoting the potential given as 
\begin{equation}\begin{split}
U(x)=&(x_1+A_1)^2(x_1-A_1)^2+(x_2+A_2)^2(x_2-A_2)^2\\&+(x_3+A_3)^2(x_3-A_3)^2,
\end{split}\end{equation}
with $\sigma=0.75.$ The variable $\xi(t)$ represents Gaussian white noise with a zero mean and a standard deviation of  $\langle\xi_i(t)\xi_j(s)\rangle=\delta_{ij}\delta(t-s)\; \forall \; i,j=1,2,3$. We choose $A_1=1.05, A_2=1.1, A_3=1.15$.

The three different choices of $A_{i}\, i=1,2,3$ generate three different heights between the potential wells, and thus three different transition rates. These rates are related to the potential through the Eyring-Kramers law~\cite{Eyring1935,Kramers1940}, which for a $d$-dimensional system states that 
\begin{equation}
    r_a^b=\frac{1}{2\pi} \sqrt{\frac{|\lambda_1(b)|\prod_{j=1}^d\lambda_j(a)}{\prod_{i=2}^d \lambda_i(b)}}e^{(U(a)-U(b))/\sigma},
\end{equation}
where $0<\lambda_1(a)<\lambda_2(a)<\dots<\lambda_d(a)$, $\lambda_1(b)<0<\lambda_2(b)<\dots<\lambda_d(b)$ are, respectively, the eigenvalues of the Hessian $\nabla^2 U(x)$ computed at the minimum of the well $a$ and at the minimum of the boundary $b$ separating the wells.

\subsubsection{The Lorenz-63 system}
We study the Lorenz-63 system \cite{Lorenz1963} given by 
\begin{equation}\begin{split}
\dot{x}(t)&=\sigma (y(t)-x(t))\\
\dot{y}(t)&=-x(t)z(t)+\rho x(t) - y(t)\\
\dot{z}(t)&= x(t)y(t)-\beta z(t),
\label{eq:lorenz63}
\end{split}\end{equation}
with $t\in[0,1000]$, and a time step $dt=0.01$. We select $\sigma=10$, $\rho=28$ and $\beta=8/3$ for which the system exhibits chaotic behavior.

\subsubsection{The stochastic Newton-Leipnik model}
We considered a stochastic version of the Newton-Leipnik model \cite{Leipnik1981}
\begin{equation}\begin{split}
\dot{x}_1(t)&=-a x_1(t)+x_2(t)+bx_2(t)x_3(t) + \sigma_1\xi_1(t),\\
\dot{x}_2(t)&=-x_1(t)-ax_2(t)+5x_1(t)x_3(t) + \sigma_2\xi_2(t),\\
\dot{x}_3(t)&=cx_3(t) - 5x_1(t)x_2(t) + \sigma_3\xi_3(t),\\
\label{eq:nl}
\end{split}\end{equation}
with $t\in[0,5000]$, $\sigma_1=\sigma_2=\sigma_3=0.04$, a time step $dt=0.01,$ and $x_1,x_2,x_3$ denoting the state variables. The parameters $a,\,b,\,c$ are positive constants, and $\xi(t)$ represents a Gaussian white-noise source with a zero mean and a standard deviation of  $\langle\xi_i(t)\xi_j(s)\rangle=\delta_{ij}\delta(t-s)$.
The Newton-Leipnik model is characterized by a double strange attractor between which the system oscillates due to the stochastic forcing.

\subsubsection{Applying the methodology}
In Figure \ref{fig:multiwell} we applied to the multi-well problem the second step of the three-step algorithm outlined above. We considered the relevant time scales of each system (via an ordering of the eigenvalues of the generator) and for each scale we display a graph showing the agglomeration of the fine clusters into connected communities as long as the corresponding modularity parameter, reported on the edge of each graph, remains positive. The resulting coarse-grained state space and graph partitioning are reported.

The variation in connectivity between different regions with the chosen time scale is evident. In particular, we notice more significant changes for those regions that are connected only through a stochastic forcing as the time scale increases. In other words, on longer time scales the transition probability between these regions becomes larger, which in turn makes them less connected. This feature can be observed for all values of the modularity parameter that correspond to bisections in the multi-well potential system. Additionally, for the Newton-Leipnik system, the top vertex of the graphs, which corresponds to the first fine-cluster bisection associated with the two different strange attractors, demonstrates the same behavior. Inside the strange attractor, it is more difficult to find similar patterns since the connected regions change across all considered time scales. It is evident from the figures that the proposed algorithm correctly groups the clusters according to the model-intrinsic time scales. For the multi-well potential, it assembles the fine clusters into eight, four, and two communities according to the three different characteristic rates of the system. The resulting coarse clusters are robust to variations in the corresponding time parameter.

For chaotic systems, the coarse clusters are more sensitive to our particular choices. In Figure \ref{fig:lorenznewton} we display our results for the Lorenz and the stochastic Newton-Leipnik systems. In the case of the Lorenz equations, phase space has been partitioned into four sets. The first branch of the modified Leicht-Newman algorithm splits state space according to the quasi-invariant sets of the Lorenz system. The second partition then further subdivides the quasi-invariant set. In the stochastic Newton-Leipnik system, the first partition of the modified Leicht-Newmann algorithm separates the two chaotic attractors. The second division then further bisects one of the chaotic attractors. 

In Figure \ref{fig:pij} we compare the elements of the transition matrix $P_{ij}(t)$ as a function of time, considering the almost-invariant sets of the Newton-Leipnik system at $\tau=1.2$ (the same value from Figure \ref{fig:lorenznewton}). The matrix elements have been constructed based on a time series of Markov states, using $Q$ from Eq. (\ref{eq_QQ}), in blue, the corrected  $Q_{\textrm{pert}}$ from our proposed method, in red, and a direct calculation of the Perron-Frobenius operator, in black. We consider the direct calculation of the Perron-Frobenius operator for time $t$ as the ``ground truth". As is apparent from the subfigures, our algorithm estimates the transition rate matrix far more accurately, considering the close match between the red and black curves. For a sufficiently long time scale, each column of the matrix converges to the steady-state equilibrium distribution, and the rows thus converge to a uniform value. 

\subsection{High-Dimensional Models}

\subsubsection{The Kuramoto–Sivashinsky equation}

\begin{figure*}
 	\centering	\includegraphics[width=1.0\textwidth]{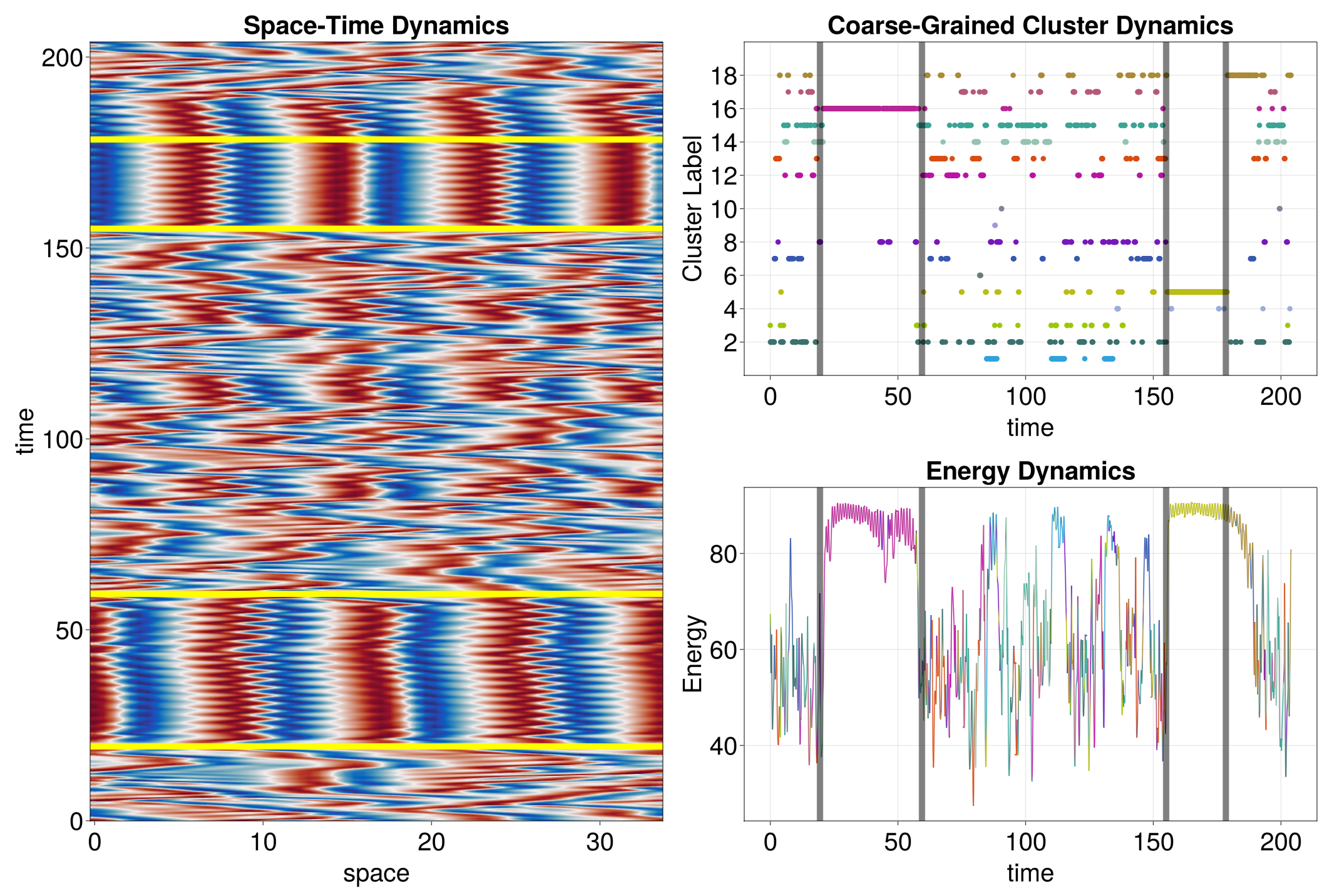}
 	\caption{Numerical solution of the Kuramoto–Sivashinsky equation on a domain of size $L = 34$ (left). We show a coarse grained projection according to states assigned by the modified Leicht-Newman algorithm using a timescale $\tau = \Delta t$ (top right). A time history of $E(t) = \int_{\Omega} u^2 dx,$ with colors associated with the coarse-grained dynamics, is shown at the bottom right. We associate the fifth and sixteenth states of the coarse grained dynamics (top right) with the regular structures and patterns observed within the two time intervals, marked by the yellow lines (left) and corresponding black lines (top and bottom right). }
 	\label{fig:ks}
\end{figure*}

Before proceeding to realistic and noise-contaminated data sets, we study the one-dimensional version of the Kuramoto–Sivashinsky (KS) equation \cite{Kuramoto1978, Sivashinsky1977, Sivashinsky1980} given as 
\begin{equation}
u_t(t,x) + u_{xx}(t,x) + u_{xxxx}(t,x) + \frac{1}{2}u_{x}^2(t,x)=0,
\label{eq:KS}
\end{equation}
which is solved on a domain of size $L = 34$ with 64 grid points. For time stepping, the nonlinear terms use a forward-Euler scheme, and the linear terms are treated by a backward-Euler method with a time step of $\Delta t = 0.017$. The system evolves to a final time of $T = 200,000$. The domain length $L=34$ has been chosen to establish chaotic transitions between two qualitatively different system behaviors: a temporally coherent solution (associated with fixed points of the underlying system), and a chaotic state.

We first use k-means with 2048 clusters where distances are defined on the entire state of 64 points. We then apply the modified Leicht-Newman algorithm with a time scale of $\tau = \Delta t$, which reduces the state space to 18 clusters. The result is displayed in Figure~\ref{fig:ks}. Among the 18 identified clusters, two are associated with the temporally coherent dynamics (in our case, states 16 and 5). The left-most plot shows a space-time plot of the Kuramoto-Sivashinsky solution, with four yellow lines serving as the start and end of time intervals during which coherent motion is observed. The top right plot shows the associated cluster labels (from the modified Leicht-Newman algorithm) as a function of time, with four black lines serving the same role as the yellow lines before. 
The bottom right plot depicts the evolution of the energy measure $E(t) = \int_{\Omega} u^2 dx,$ using the same coloring scheme as the coarse-cluster dynamics in the top right figure. A clear correlation between the energy dynamics and the temporally coherent structures is discernible. 

\subsubsection{Experimental data from PIV measurements}

\begin{figure*}
 	\centering	\includegraphics[width=0.5\textwidth]{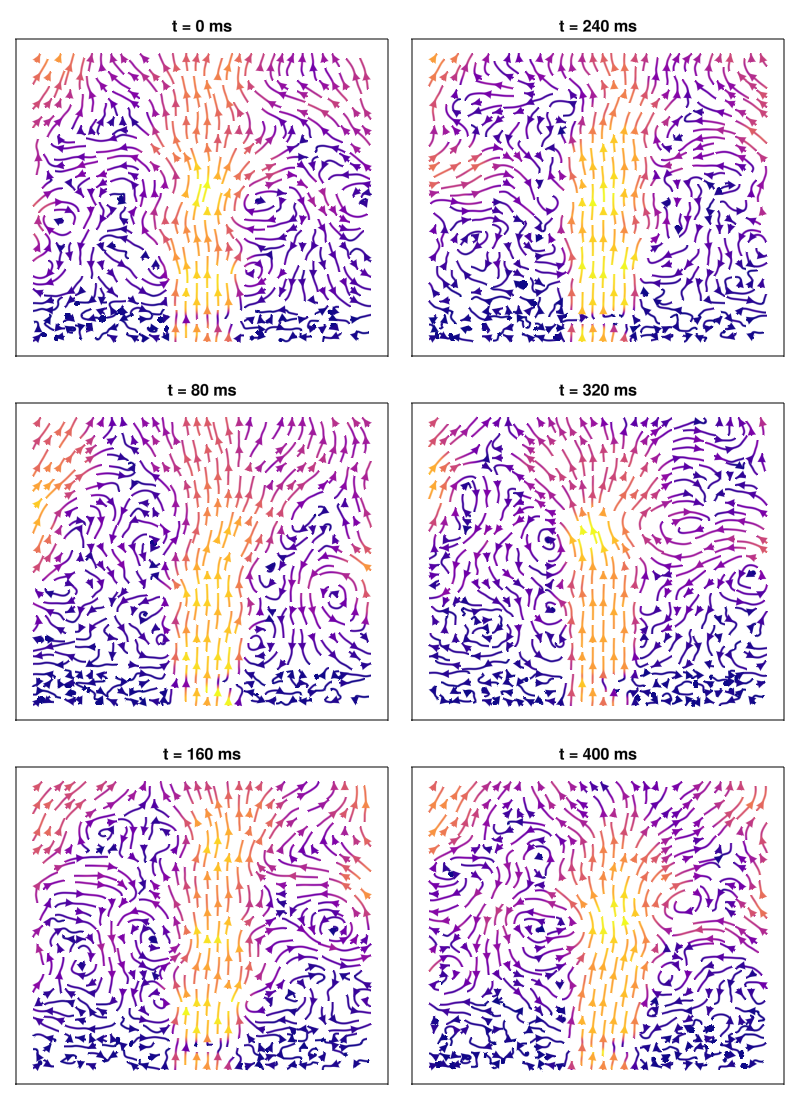}
 	\caption{First eight snapshots of the particle-image velocimetry data with the corresponding time. The color of the arrows corresponds to the speed of the velocity field, with lighter colors corresponding to faster speeds. }
 	\label{fig:piv}
\end{figure*} 

\begin{figure*}
 	\centering	\includegraphics[width=0.95\textwidth]{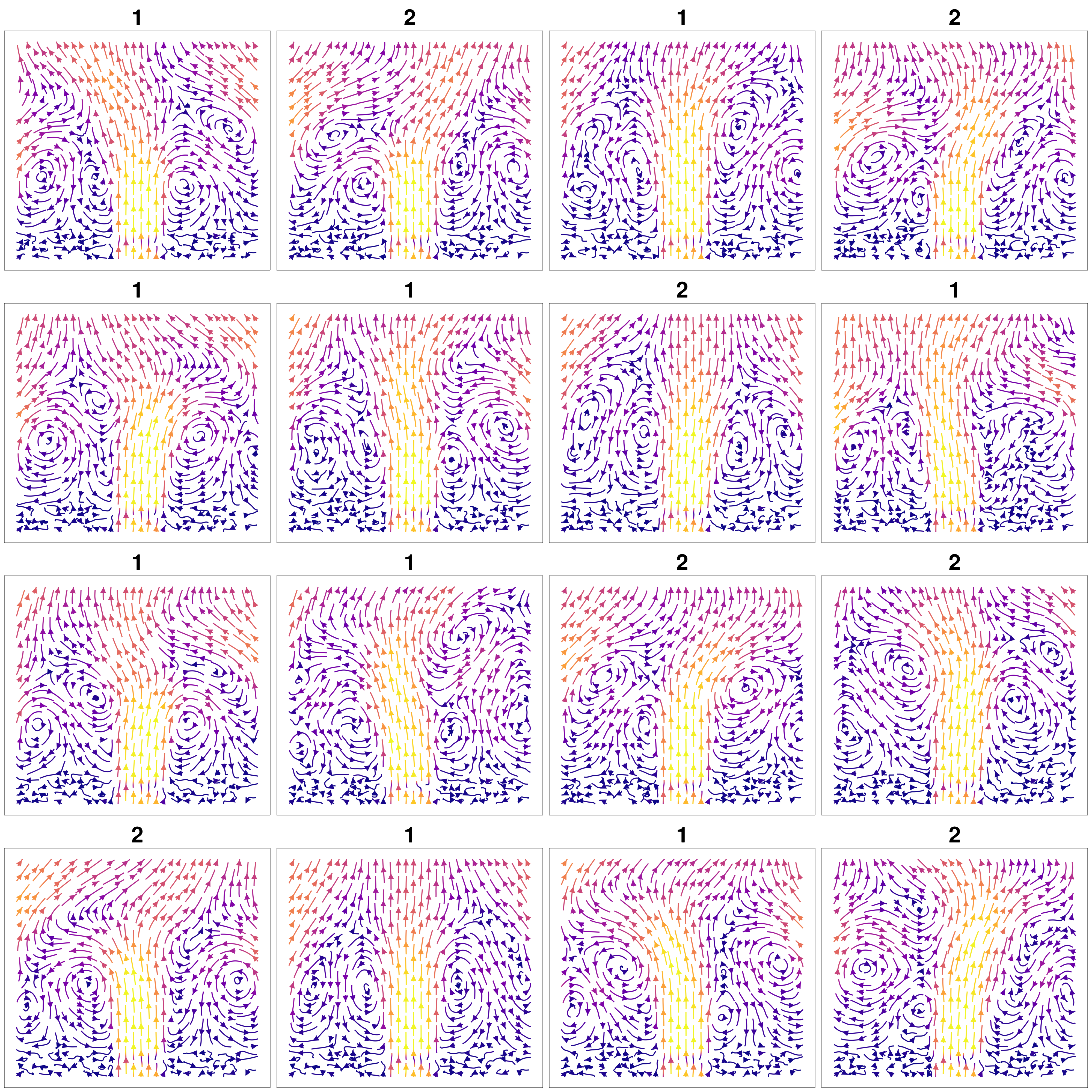}
 	\caption{Clustering of PIV data sequence. First, an SVD is taken of the data and only the first 100 modes are retained. Then, the remaining time series is clustered into 16 groups using a k-means algorithm. Finally, the modified Leicht-Newman algorithm is applied, revealing two main partitions of state space corresponding to a flow that sways between a predominantly left and right state. }
 	\label{fig:piv_cluster}
\end{figure*} 

The final example consists of data sampled from experiments via Particle-Image Velocimetry (PIV). The data set describes the transverse flow through a cylinder bundle, as encountered in various industrial applications such as, e.g., cooling rods or heat exchangers. The cooling fluid, emerging from the cylinder bundle, forms a jet that quickly becomes unstable and settles into a quasi-periodic limit-cycle behavior with two attracting dynamics, referred to as a `flip-flop state.' Similar quasi-bistable phenomena occur, for example, in the wake past bluff bodies. 

In our case, we consider a time-resolved data sequence of two-dimensional velocity-field slices in an interrogation domain of $40.36 \textrm{mm}$ in the streamwise (vertical) and $32.08 \textrm{mm}$ in the cross-stream (horizontal) direction, which is discretized into a $63 \times 79$ Cartesian and equispaced grid. Only two in-plane velocity components are recorded. The data sequence is sampled uniformly in time, with a $4 \textrm{ms}$ distance between two consecutive snapshots. With a cylinder gap of $10.7 \textrm{mm}$ and a mean jet velocity of $0.663 \textrm{m/s},$ the resulting Reynolds number based on the volume flux (18 $\textrm{m}^3/\textrm{h}$) and the cylinder diameter ($12 \textrm{mm}$) comes to $Re=3000.$ 

The jet progresses through a sequence of flow-transverse oscillations with two quasi-equilibrium points emerging: a left- and right-leaning mean state about which the jet fluctuates. In contrast to previous examples, the processed data set contains a considerable amount of measurement noise which will probe the robustness of the clustering algorithm and the subsequent data analysis.

In Figure \ref{fig:piv}, we plot eight snapshots of the PIV data. The dimensionality of the data has been reduced by considering only the first 100 singular vectors, which account for 93$\%$ of the total variance. 

Figure \ref{fig:piv_cluster} shows the centers of the eight clusters used in the k-means algorithm to divide the state space. In addition, we display the semi-invariant sets assigned to each of cluster obtained from the modified Leicht-Newman algorithm with a time parameter of $t=4\textrm{ms}$. The algorithm correctly separates the left- from the right-leaning mean state about which the jet fluctuates. 

\section{Conclusions}

In conclusion, we have presented two algorithms that are used in tandem to produce a reduced statistical description of a dynamical system. The first algorithm focused on determining an optimal partition of state space based on a new modularity criteria adapted from the Leicht-Newman algorithm. The second algorithm determined a data-driven generator for a stochastic process that works over a broad range of temporal scales. We then applied the methodology to five different systems: three low-dimensional dynamical systems and two high-dimensional dynamical systems. The low-dimensional systems were selected to demonstrate the method on stochastic and chaotic dynamical systems. The high-dimensional systems were chosen to illustrate the ability of the method to provide information on the underlying system.  

There are various future directions for the present study. Applying the methodology to other dynamical systems is expected to yield insight into the structure and the statistical properties of the system. In particular it would be interesting to apply the methodology to high-dimensional dynamical systems arising from fluid mechanics or climate science \cite{Froyland2018, gosha_2024}. Furthermore, as mentioned in the introduction, modularity maximization has a number of known problems \cite{PhysRevE.81.046106, Fortunato_2007, Fortunato_2016, Peixoto_2023}. Developing new clustering algorithms for Perron-Frobenius operators is a promising direction for further study. 

\section*{Acknowledgements}
The authors want to thank the 2022 Geophysical Fluid Dynamics Program, where a significant portion of this research was undertaken which is supported by the National Science Foundation, United States and the Office of Naval Research, United States. L. T. Giorgini gratefully acknowledge support from the Swedish Research Council (Vetenskapsrådet) Grant No. 638-2013-9243. Nordita is partially supported by Nordforsk. A. N. Souza is supported by the generosity of Eric and Wendy Schmidt by recommendation of the Schmidt Futures program. 

 \bibliographystyle{elsarticle-num} 
 \bibliography{cas-refs}





% \end{thebibliography}
\end{document}